# A possible fundamental theory of Cooper Pair formation in unconventional superconductivity

Godfrey E. Akpojotor[1] and Alexander E. Animalu[2,3]
[1]Department of Physics, Delta State University, Abraka 331001, Nigeria
[2]Department of Physics and Astronomy, University of Nigeria, Nsukka, Nigeria
[3]International Centre for Basic Research, 20, Limpopo Street, FHA, Maitama Abuja, Nigeria
E-mails: [1]akpogea@delsung.net and [2]ibr32@aol.com

**Abstract**
The successful application of the electron-phonon interaction (EPI) mechanism in formulating the Bardeen-Cooper-Schrieffer (BCS) theory of superconductivity is among the most outstanding intellectual achievements in theoretical physics because of the successful application of the theory to the conventional superconducting materials. Therefore, its unsuccessful application to the nonconventional superconducting materials has led to the search for new theories which generalized it, include an interplay with other mechanisms or are formulated from non – EPI mechanisms. We observe in this current study that to achieve a generalized theory of superconductivity, there is need to first developed a quantitative structure model of the Cooper pair formation (CPF) in line with the formulation of the molecules in nature which has given birth to hadronic mechanics. This generalized formulation is the iso-superonductivity model which is based on the observation that the Cooper pair of the standard BCS model may have a nonlocal-nonhamiltonian structure equivalent to the strong interaction ("hadronic" mechanics (HM)) structure of the neutral pion, as compressed positronium atom at short distances (< 1 F ~ $10^{-13}$ cm. The equivalent but approximate description at large distances (> 1 F) is by the quantum mechanical superexchange interaction. This generalized CPF has been used to successfully explain the high-$T_c$ superconductivity in the cuprates as well as to account for the transition temperatures of these materials. It is also used to account for the high-$T_c$ iron based superconducting materials.

**PACS:** 74.20._z, 74.62.Bf; 75.30.Et,

## Introduction

The year, 2011, was declared by the international scientific community as the centenary celebration of the discovery of superconductivity (SC) in Hg at about 4 K by the Dutch physicist, Heike kamerlingh Onnes on April 8, 1911. Today, there is metallic/non-metallic, conventional/nonconventional, doped/undoped and low/high $T_c$ superconducting materials at ambient/non-ambient pressure [1]. The only generally accepted theory [2] is that by Bardeen-Cooper-Schrieffer (BCS) in 1957 and its salient feature is the electron-phonon interaction (EPI) leading to the formation of an ensemble of Cooper pairs which can propagate coherently as the superconducting state of the parent material. The BCS theory has been used to account for a number of metallic and intermetallic superconductivity and all such materials are known as conventional superconductors and their transition temperatures are low [3,4]. This BCS theory has led to a number of important applications like superconducting magnets for laboratory use (spanning from small-scale laboratory experiments to the large hadron collider (LHC) 's bending magnets), and importantly, for magnetic resonance imaging (MRI) systems as well as explanation of puzzling experimental data such as nuclear magnetic resonance (NMR) relaxation rate and Josephson tunneling [5]. However, the BCS failed to explain the Meissner effect which is a fundamental property of superconductors [6] and has been proven incapable of predicting high – temperature superconductors [7] and providing the guidelines to search for new materials [8]. Therefore the BCS theory did not only fail to predict the relatively high $T_c$ of the superconducting copper oxide compounds commonly known as cuprates discovered in 1986 [9] but also failed to account for this class of superconductors [10 - 13]. In general, the BCS has failed in its application to a number of new classes of superconductors discovered since 1970 [6]. To account for these superconductors now collectively known as non-conventional superconducting materials, there has been a deluge of proposals of new theories which generalized the BCS theory by replacing the phonon with other bosons [14 – 16], introduce an interplay of the EPI and other mechanisms [17 - 19] or are formulated from non – EPI mechanisms [20 - 25] as well as those that even question the validity of the BCS theory [6, 26]. However, the partial success of these theories for the superconducting materials they are developed for, means the main properties of the superconducting phenomena are still poorly understood and new concepts are needed [27 - 29]. The basic requirement of any theory of superconductivity is to first provide a mechanism of the Cooper pair formation (CPF) and its coherent propagation. Therefore the basic question man has been struggling with is how does two electrons which should repel as provided by quantum mechanics come to bind to form the Cooper pair? A general answer to this question can be achieved by providing a quantitative and qualitative structure model of the CPF possible in all superconducting materials. In this current study, we have shown that at short distances, the CPF formation is by a nonlinear, nonlocal and nonhamiltonian strong hadronic-type interactions due to deep wave-overlapping of spinning particles leading to Hulthen potential that is attractive between two electrons in singlet couplings [20, 21, 29] while at large distance the CPF is by superexchange interaction which is purely a quantum mechanical affairs [23, 26, 30]. We observe that for both distances the control parameter responsible for the superconducting state is the effective valence z. Therefore $T_c$ expressions depending on the effective valence were obtained for both the superconducting cuprates and the iron based compounds to verify experimental results.

## Brief review of the EPI mechanism of CPF

The mechanism of electron-electron interaction mediated by phonon emanates from the discovery of the isotopic effect in superconductivity [31] which suggested that the vibration of the lattice could be involved in the interaction. Usually, electrons are scattered from one state k to another state K' leading to electrical resistivity. The lattice vibrations which distort the local crystal structure and consequently the local band structure can be absorbed and emitted in the scattering process. The first electron, which absorbed the phonon, has its effective mass increased and consequently will attract it to a second electron thereby leading to the CPF. Together, all the paired electrons form a condensate that moves as a single entity resulting in the zero resistance. This proposal was boosted by the earlier observation that the $T_c$ of superconductors depends on the isotopic mass of the lattice, in the independent studies of Reynolds et al. [32] and Maxwel [33]. Therefore one of the key features of the emanating BCS theory is its $T_c$ expression,

$$T_c \approx \frac{\hbar \omega}{k_\beta} e^{-1/g}, \qquad (1a)$$

where $\omega$ is the relevant or characteristic phonon frequency, $\hbar$ is the Plank's constant and $k_\beta$ is the Boltzman's constant while the coupling factor g is define by

$$g = \lambda + \mu^* \approx N(E_F)V = \lambda, \qquad (1b)$$

where $\lambda$ is the EPI coupling constant, $N(E_F)$ is the density of state at the Fermi level and V is the EPI coupling strength. Note that the effect of the Coulomb repulsive parameter $\mu^*$ is considered negligible, that is, the V is completely dominant.

It is easy to observe in Eq. (1a) that when the coupling constant $\lambda$ is small and the value of $k_\beta T_c$ is of the same order of magnitude as the binding energy of a Cooper pair, at $T > T_c$, the Cooper pairs break apart and the material returns to normal state.

As expected from a successful theory, the BCS $T_c$ formula has been used to obtain the $T_c$ of many conventional superconductors that agree fairly with experiments. The $T_c$ expression however, has some shortcomings such as its inability to account for the deviation of some superconductors from the isotopic shift predicted in the study of Ref. [38]: the property of isotopic effect in superconductivity is that the $T_c$ is proportional to the isotopic mass, that is,

$$T_c \propto M^{-\beta}. \qquad (2)$$

where M is the isotopic mass and $\beta$ is the isotopic shift. It is obvious in Eq. (1a) that the $T_c$ is proportional to $\omega$. The $\omega$ however depends on the ionic mass M, that is

$$\omega = (k/M)^{1/2} \qquad (3)$$

where k is a constant.

The implication is that two isotopes of the same superconducting materials have different critical temperatures because as shown by Eqs (2) and (3), they have different bandwidths or frequencies. The β observed for most non-transition metals is about 0.5 while for transition metals, the values are much smaller than the BCS predicted values and nearly absent in some metals like Zr and Ru (β = 0.00 + 0.05). Similarly, the BCS $T_c$ expression could not account for why the gap ratio of some metals deviate from the constant value, i.e. $\Delta_0(0)/K_\beta T_c = 3.52$, predicted for them.

In Eq. (1a), the $T_c$ obviously depends on two factors: $\omega_{ph}$ and λ. An increase in any of these parameters will increase the $T_c$ value. However, the basic assumption as stated above, requires that the λ should be small. This implies that the $T_c$ cannot be increased by arbitrarily increasing the λ. It is this restriction of the value of the λ to small values that makes the BCS theory to be applicable only to weak-coupling superconductivity and hence the limitation of its predicting power to $T_c \leq 25 K$ [7]. As stated above, to generalize the BCS theory, there have been various attempts in the literature such extension of the EPI to strong coupling regime resulting in the Migdal- Eliashberg Theory [34, 25], polaron formation believed to occur beyond the maximum coupling limit of a normal EPI [36], replacement of the phonons with higher energy bosons (like excitron or charge fluctuations [14], plasmons [15], magnons [16], etc), including the effect of Coulomb repulsion to the original BCS interactive potential [37], including other bands [38], etc. All these attempts have not recorded generally accepted success as expected in their application beyond the conventional superconductors.

**A possible structure model of the CPF in all superconducting materials**

A simplistic overview of the emergence of superconducting materials since 1970 for which there is a consensus that they cannot be described by the EPI mechanism of the BCS theory or at least there are serious doubts whether they can, identify at least ten distinct materials or families of materials [6]: (i) the cuprates, hole-doped ($YBa_2Cu_3O_7$) and electron-doped ($Nd_{1-x}Ce0_xCuO_{4-y}$); (ii) heavy fermion materials ($CeCu_2Si_2$, $UBe_{13}$, $UPt_3$); (iii) organics ($TMTSF_2PF_6$); (iv) strontium–ruthenate ($Sr_2RuO_4$); (v) fullerenes ($K_3C_{60}$, $Cs_3C_{60}$); (vi) borocarbides ($LuNi_2B_2C$, $YPd_2B_2C$); (vii) bismuthates ($Ba_{1-x}K_xBiO_3$, $BaPb_{1-x}Bi_xO_3$); (viii) 'almost' heavy fermions ($U_6Fe$, $URu_2Si_2$, $UPd_2Al_3$); (ix) iron arsenide compounds ($LaFeAsO_{1-x}F_x$, $La_{1-x}Sr_xFeAs$); (x) ferromagnetic superconductors ($UGe_2$, $URhGe_2$). It is pertinent to mention the recent prediction of superconductivity in neutron stars [38, 39], alcoholic beverages induce superconductivity in $FeTe_{1-x}S_x$ [40] as well as the possibility of superconductivity in a vacuum [41]. The theoretical proposals to account for the superconductivity in these materials are based on their possible structures and the possible means of naturally achieving the CPF from the appropriate bonding of the elements yielding these structures. The first quantitative representation in line with this thinking emanates from the Santilli's proposal in 1978 [29, 42] to build the foundation of hadronic mechanics wherein a bound state of one electron and one positron at a short distance (< 1 F ~ $10^{-13}$ cm) with non-local, non-linear and non-potential is due to deep overlapping of their wavepackets. Animalu observed that at such distances, the magnetically induced Hulthen potential which is an attractive force will dominate the Coulomb repulsion between two electrons to allow them to bond into singlet coupling as in the CPF in the cuprates [20, 21, 25]. At large distances (> 1 F ), the Hulthen potential no longer dominates and it has been suggested by Akpojotor [25] that the CPF is by superexchange interaction which naturally affects electrons that are close enough to have (no deep) overlapping wavefunctionis and this is purely a quantum mechanical affairs [29]. This dichotomy emanates from the regimes of validity of hadronic mechanic and quantum mechanics. For sufficiently large distances, particles can be described effectively in point-like approximation so that there is the sole presence of action-at-a distance potential interaction which can be represented by a Hamiltonian. This approximate point-like description is no longer valid at sufficiently small distances so that their interaction is dominated by a contact non-potential character. As explained in Ref. [29], the quantum mechanical point-like description is only not applicable in this regime since it is not violated because the condition in the regime was beyond what quantum mechanics was conceived for. Thus the possible structure model of the CPF emanates from the strong valence at short distances of hadronic mechanics while at large distances the equivalent mechanism is by the quantum mechanical superexchange interaction. This is the foundation of the isosuperconductivity model of superconductivity.

It has been suggested in Refs. [20, 21, 25] that the control parameter to drive these interactions which is known in hadronic mechanics as the 'trigger' is the effective valence. Here the 'trigger' assumes the same meaning as the often ambiguous 'under favourable condition' that is required for the EPI to lead to the CPF in the BCS theory. The achievement of the effective valence as the natural favourable condition for the CPF by either the hadronic mechanical strong valence or quantum mechanical superexchange interaction is in line with the common knowledge that in material design, the valence electrons are known to govern the crystal structure [43, 44].

**Application of the iso-superconductivity model**

The two classes of superconducting materials with relatively high transition temperatures than the others are the superconducting cuprates and iron based compounds. Consequently these two classes of materials have received the highest interest of researchers in the field of superconductivity which can be grouped into those seeking for a theory to account for these materials and those seeking for how to increase their $T_c$ to room temperature. Thus our application here of the isosuperconductivity model will be restricted to these two classes of materials. In particular, it is straightforward to apply the model to the superconducting cuprates since they have the $CuO_2$ planes common to all members of the family as the key to understanding them. Further, the superconductivity is driven by the effective valence of the Cu so that a $T_c$ that depends on this z is obtained. The iron based compound, however, do not have a common plane: the consensus feature of members of this family is that the Fe is responsible for their high $T_c$ [43, 45]. Thus a $T_c$ that depends on this z will also be obtained for the superconducting iron based compounds.

*The superexchange interaction in the $CuO_2$ planes of the superconducting cuprates*

One early consensus after the discovery of the high $T_c$ superconducting cuprates, is that the key to understanding these materials is the $CuO_2$ planes common to all of them. Another general consensus is the Anderson observation that the superconductivity phenomenon in the cuprates involving the Cu 3d and O 2p bands of the $CuO_2$ plane can be reduced to an effective single band pairing problem via his doped resonant valence bond (RVB) model [48]. Zhang and Rice generalized the Anderson approach by showing that it is possible to obtain the Cooper pair as a bound singlet state of the $CuO_2$ plane within a t-J model [49]. This singlet state now commonly known as the Zhang-Rice singlet (ZRS) which is the Cooper pair of the superconducting cuprates has been achieved [23] by the quantum mechanical superexchange interaction using the first electron removal (FER) approach [50] and a highly simplified correlated variational approach [51] to obtain a t-J model with the interaction part, J being the XY limit of the anisotropic Heisenberg exchange interaction in second quantization language:

$$H_{t_{pd}} = -t_{pd}\left[\sum_{\{i\}}\sum_{<j,k>\in\{i\}} d^+_{i_{r'},\sigma} p^+_{j_{r'},\sigma} d_{i_r,\sigma} p_{k_r,\sigma} + H.C.\right] +$$

$$J_{dp}\left[\sum_{\{i\}}\sum_{<j,k>\in\{i\}} d^+_{i_{r'},\sigma} p^+_{j_{r'},\bar{\sigma}} d_{i_r,\bar{\sigma}} p_{k_r,\sigma}\right] \qquad (4)$$

where $d^+(d)$ is the creation (annihilation) of the carrier at the Cu $3dx^2-y^2$ orbital and $p^+(p)$ is the creation (annihilation) of the carrier at the O $2p_x$ and O $2p_y$ orbitals and the $t_{pd}$ denotes a hopping between a Cu and O in the same plane.

This model gives the experimental bandwidth of $W \approx 1$ eV [52] for the common value in the literature ($J_{dp}/t_{pd} = 0.3$) [53]. One important evidence supporting our mechanism of Cooper pairs formation and their propagation is the observation in [54] that the critical $T_c$ of the superconducting cuprates depends on the number of $CuO_2$ planes within a short distance of each other in the structure. The implication is that the smaller the $CuO_2$ planes, the more the enhancement of the hybridization of the carriers in the Cu and O sites into the ZRS due to decreasing distances of their overlapping wavefunctions as expected from the structure mode of the CPF in this current study. Finally we re-emphasize that the description in this section is valid as the quantum mechanical approximation of the hadronic mechanical reality, to which we now turn to.

*The strong valence CPF and* **the $T_c$ expression for** *the superconducting cuprates*

The strong valence hadronic mechanical CPF began with the observation by Animalu [20 - 21] that for the strongly correlated high-$T_c$ cuprate materials, $(...)Cu_m O_{n-x}$, the Santilli representation of the neutral pion as a compressed positronium system ($e^+ e^-$) system [42] is equivalent to a state of mutual overlap/non-orthogonality of the paired Cu 3d and O 2p electrons wavefunctions, $\psi_\uparrow$ and $\psi_\downarrow$ such that $<\psi_\uparrow / \psi_\downarrow> \neq 0$.

By iso-unitary transformation in line with the Santilli Lie-isotopic/Lie admissible approach (29), Animalu was able to transform the BCS theory into a t-J model for the superconducting cuprates [20, 21, 25]. Starting with the Lurie-Cremer [55] quasiparticle wave equation,

$$i\frac{\partial}{\partial t}\Psi(\mathbf{r},t) = H\Psi(\mathbf{r},t), \qquad H \equiv \frac{1}{2m}\mathbf{p}^2 \tau_3 + \Delta \tau_1 \qquad (5)$$

via the non-unitary ("isotopic lifting") transformation of the underlying "metric" ($g$),

$$g \equiv \tau_3 = \begin{pmatrix} 1 & 0 \\ 0 & -1 \end{pmatrix} \rightarrow \hat{\tau}_3 = \begin{pmatrix} 1 & 0 \\ 0 & -T \end{pmatrix} \equiv g \qquad (6)$$

which is characterized by a nonlocal integral (pseudopotential) operator defined by

$$T\psi_\downarrow^*(\mathbf{r}) = \int d^3 r' [\delta(r-r') - \psi_\uparrow^*(\mathbf{r})\psi_\uparrow(\mathbf{r}')]\psi_\downarrow^*(\mathbf{r}') \qquad (7)$$

where $\psi_\uparrow^*(\mathbf{r})$ and $\psi_\downarrow(\mathbf{r})$ are the two spinor components of the quasi-particle wavefunction $\Psi(\mathbf{r},0)$ in the Nambu representation, $p^2/2m$ being the kinetic energy operator (measured from the Fermi level) and $\Delta$ is the pair potential energy. It is apparent from Eq. (7) that when the overlap integrals or "orthogonalization term"

$$Z^{\frac{1}{2}} = \int d^3 r' \psi_\downarrow^*(\mathbf{r}')\psi_\uparrow(\mathbf{r}') \equiv \langle \psi_\downarrow^* | \psi_\uparrow \rangle \qquad (8)$$

is zero, $T$ reduces to unity and we recover the standard (BCS) model exactly. Since we may rewrite $T$ in the form

$$T = 1 - |\psi_\downarrow\rangle\langle\psi_\uparrow^*| \qquad (9a)$$

so that $T^2 = T$ if $\langle\psi_\downarrow^*|\psi_\uparrow\rangle \neq 0$, the physical effect of $T$ is that the charge on the $e^- \uparrow$ representated by the expectation value of T, i.e.,

$$\langle\psi_\downarrow^*|T|\psi_\downarrow\rangle = 1 - Z \qquad (9b)$$

is "depleted" by an amount Z (called the "orthogonalization charge") whereas the charge on $e^- \downarrow$ appears to vanish, i.e.,

$$\langle\psi_\uparrow^*|T|\psi_\uparrow\rangle = 0 \qquad (9c)$$

In other words, $e^- \uparrow$ behaves like a neutral spin-$\frac{1}{2}$ quasiparticle (spinion) while $e^- \downarrow$ behaves like a fractionally-charged quasiparticle ("anyon"). Consequently, in the solid state where the wavefunction $\psi_\sigma(r,t)$ to which the nonlocal transformation in Eq.(6) is to be applied is related to the $\phi_i(r)$ and $c_{k\sigma}(t)$ of the second quantized formulation by

$$\psi_\sigma(r,t) = \sum c_{k\sigma}(t)\phi_i(r) \qquad (10)$$

the corresponding transformation of the corresponding creation and annihilation operators, $c_{k\sigma}^+$ and $c_{k\sigma}$, into iso-creation and iso-annihilation operators, is defined by

$$\hat{c}_{ik\sigma}^+ = T_{ik\sigma}c_{ik\sigma}^+ \equiv (1-n_{ik\sigma}), \quad n_{ik\sigma} = c_{ik\sigma}^+ c_{ik\sigma} \qquad (11)$$

and similarly for $\hat{c}_{ik\bar{\sigma}}$ where $\sigma = \uparrow$ for $\bar{\sigma} = \downarrow$ and vice versa. This has the effect of transforming the hopping (kinetic energy) term *exactly* into

$$-t\sum_{\langle ij\rangle,\sigma}\hat{c}_{i\sigma}^+\hat{c}_{j\sigma} = -t\sum_{\langle ij\rangle,\sigma}(1-n_{i\sigma})c_{i\sigma}^+c_{j\sigma}(1-n_{j\bar{\sigma}}). \qquad (12)$$

as in Eq.(4) characterizing the $t - J$ model. It follows that the difference between the $t - J$ model and the isosuperconductivity model lies in the replacement of the U-term in the Hubbard model by the J−term in the t − J model (with the antiferromagnetic exchange constant $J = t^2/U$ via second-order perturbation theory). Typically, $i(j) = d, p$ label electrons (bands) of Cu 3d and/or O 2p characters whose wavefunctions may overlap and/or bands hybridize; and (i(j) = 1, 2, ...,N) in the nearest-neighbour electron transfer (hopping) integral.

By virtue of the transformation defined by Eq.(11), only single occupancy per spin site is permitted but double occupancy of an orbital site is not forbidden. Another feature of the second-quantized theory form of the iso-creation and iso-annihilation operators is that the waveoverlapping is associated with the coexistence of a non-zero antiferromagnetic spin wave state, $\langle c_{i\downarrow}^+ c_{i\uparrow}\rangle \neq 0$ and Cooper pair state $\langle c_{i\uparrow} c_{i\downarrow}\rangle \neq 0$ under Gor'kov's factorization of the products of three fermion creation and annihilation operators involved in the transformation

$$\begin{aligned}T_{ij}c_{i\downarrow} &= (1-c_{i\uparrow}^+c_{i\uparrow})c_{i\downarrow}\\ &\equiv c_{i\uparrow}c_{i\uparrow}^+c_{i\downarrow} \rightarrow \langle c_{i\uparrow}c_{i\uparrow}^+\rangle c_{i\downarrow} - \langle c_{i\uparrow}c_{i\downarrow}\rangle c_{i\uparrow}^+ + \langle c_{i\uparrow}^+c_{i\downarrow}\rangle c_{i\downarrow}\end{aligned} \qquad (13)$$

In this (mean field) sense, one can derive from the isosuperconductivity model one of the primary objectives of the $t – J$ model which is to describe the coexistence of superconductivity and antiferromagnetism in high-$T_C$ materials as a function of band filling. Thus we have shown here the CPF by either the hadronic mechanical strong valence or quantum mechanical superexchange interaction can be used to formulate a t-J model that can be used to account for the formation and propagation of the ZRS in the superconducting cuprates. The most important difference between the standard t−J model and the iso-superconductivity model lies in the ability of the latter to predict $T_c$ from an exact solution of the model. Another important feature of the isosuperconductivity model is that instead of solving an integral equation for the energy gap as done in the conventional BCS model, the desired result comes from the self-consistent solution of the conventional Schrodinger equation for one spin state,( $\psi_\downarrow$ ) say,

$$H\psi_\downarrow \equiv (p^2/2m + V_C)\psi_\downarrow = E_\downarrow\psi_\downarrow, \qquad (14a)$$

in the Coulomb field $V_C$ of the $Cu^{z+}$ ion "trigger" in Figure. 1b, and an iso-Schrodinger equation

$$HT\psi_\uparrow \equiv (p^2/2m + V_H)\psi_\uparrow = E_\uparrow\psi_\uparrow, \qquad (14b)$$

for the opposite spin state($\psi_\uparrow$), where T is the non-local (psuedopotential) integral operator defined by Eq.(7). This has the effect of replacing the Coulomb potential, $V_C$, by an effective Hulthen potential, $V_H$ in Eq.(14b) for the Zhang-Rice singlet (e− ↓, e− ↑):

$$V_C \to V_C - \frac{E_\downarrow \langle \psi_\downarrow | \psi_\uparrow \rangle \psi_\downarrow}{\psi_\uparrow} = -V_0 \frac{1}{e^{kr} - 1} \equiv V_H \qquad (15)$$

where $V_0$ is proportional to $\langle \psi_\downarrow | \psi_\uparrow \rangle$. From the exact solution of Eq.(14b), Animalu [20, 21] derived the following formula for the critical temperature having the general form:

$$T_C = \frac{\Theta_J}{\left[\exp(\frac{1}{NV}) - 1\right]} \qquad (16a)$$

where *NV* reprsents the dimensionless coupling constant while

$$\Theta_J = \frac{\hbar \omega_p}{k_B \sqrt{d\varepsilon(q_D)}} \qquad (16b)$$

is the "jellium" temperature, *d = 1, 2, 3* being the effective dimensionality of the system and $\varepsilon(q_D)$ the Hatree dielectric function evaluated at the Debye wavenumber $q_D$. We observe that in the weak coupling limit *NV < 1*, we may express the result in the BCS form:

$$T_C = \Theta_J \exp(-1/NV) \qquad (17a)$$

But in the strong coupling limit, i.e. if *NV > 1*, we may expand the exponential in the denominator of Eq.(2.16a) to first order in *1/NV* to get

$$T_C = \Theta_J NV \qquad (17b)$$

Solving the explicit form of Eq.(17a) used in Ref. 21 for the verification with experimental data in the cuprates with structural formula $(...)Cu_m O_{n-x}$ yields:

$$T_C = \Theta_J \exp(-\frac{13.6}{z}) = \left(\frac{367.3z}{\sqrt{d\varepsilon}}\right) \exp\left(-\frac{13.6}{z}\right) \; (^0K) \qquad (18)$$

where the effective valence z of the $Cu^{z+}$ ion is given by $z = \frac{2(n-x)}{m}$.

Eq. (18) emphasizes the foundation of the iso-superconductivity model for the superconducting cuprate materials, $(...)Cu_m O_{n-x}$, that a $Cu^{z+}$ ion of effective valence, z ≡ *2(n − x)/m* provides a "trigger" for the overlapping (i.e., "covalent" mixing) of electron wavefunctions to form a singlet pair, (e−↓, e−↑)$_{HM}$. (see Figure 1). The results of Eq. (18) which have been published in earlier works [20, 21, 25] and compared with experimental results are shown in Table 1 and Figure 2.

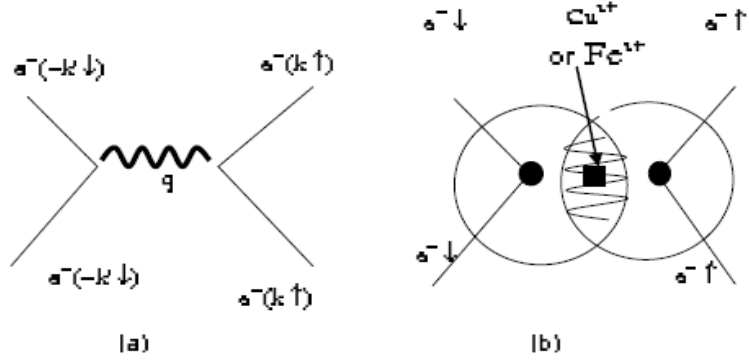

**FIGURE 1.** (a) Attractive electron-electron interaction mediated by virtual phonon exchange in the conventional BCS model; (b) attractive electron-electron pairing due to deep overlapping of electron wavefunctions around $Cu^{z+}$ or $Fe^{z+}$ ion "trigger" in orbital $s(\uparrow,\downarrow)$ state envisaged in iso-superconductivity model.

*The Cooper pair formation and $T_c$ prediction in the iron based superconducting materials*

In January 2008, the group of Hideo Hosono in Japan reported that a layered iron arsenide material (LaOFeAs) is superconducting with a transition temperature ($T_c$) of 26 K [56]. The two surprising issues here are: (1) it was believed before then that iron ions in many compounds have magnetic moments and consequently form an ordered magnetic state [57] rather than a superconducting state though the Hosono group had earlier obtained a $T_c$ of 5 K in LaOFeP [58]; (2) The $T_c$ of 26 K is higher than any intermetallic compounds except that of 35 K in $MgB_2$ [59]. Thus it has attracted a deluge of experimental and theoretical studies of the iron based compounds commonly called the iron pnictides (pnictides means compounds of the nitrogen group). Today, a $T_c$ of 55 K has been achieved [60] and there are four major classes of this family of superconducting materials: iron oxypnictides/single layered LnOMPn (Ln = La, Ce, Pr, Nd, Sm, Gd, Tb, Dy, Ho and Y: M = Mn, Fe, Co and Ni: Pn = P and As) ; oxyfree-pnictides/single layered AMPn (A = LnO = Li and Na: M = Mn, Fe, Co and Ni: Pn = P and As); oxyfree-pnictides/double layered ALM2Pn2 (AL = Ba, Sr, Ca: M = Mn, Fe, Co and Ni: Pn = P and As) and chalcogen/non-layered MCn (M = Mn, Fe, Co and Ni: Cn = S, Se and Te).

It has been observed that the superconducting $T_c$ of the iron based compound is critically dependent on extremely small changes in the iron stoichiometry [45]. For example, the non-iron based superconductor, LaONiP is isostructural to LaOFeAs, yet it can be accounted for by electron-phonon coupling [61] while the electron-phonon interaction (EPI) of the iron-based compounds has been shown to be too weak to produce their remarkably high $T_c$ [62 - 63]. This observation led to the speculation that the superconductivity of the iron-based compounds might be related to that of the cuprates. The speculation was boosted by early observation that like the cuprates, the parent compounds of the iron pnictides are antiferromagnetic (AF) and only become superconducting when doped [64 - 65]. Further, just as the three bands problem of the $CuO_2$ planes common to the cuprates can be reduced to an effective single band pairing problem via the Anderson doped resonant valence bond model and its generalization by Zhang and Rice into the t – J model, so also the high-$T_c$ superconductivity in the iron-based compounds known to involve multi-orbital effects of the Fe-3d with filling of approximately six electrons per Fe-site in the pnictides has been shown in the 2009 selfconsistent fluctuation exchange (FLEX) model by Zhang et al [66] to be reducible to an orbital $s(\uparrow, \downarrow)$ coupling affair also known as the $s_\pm$ state [67]. It has been shown that such an $s_\pm$ state can be achieved as the CPF from superexchange interaction within a quantum mechanical treatment [68 - 69]. The analogy for the natural description of the CPF within hadronic mechanics is similar to the situation in the foundation of the iso-superconductivity model for the high-$T_c$ cuprate materials: here an $Fe^{z+}$ ion of effective valence of appropriate value provides a "trigger" for the deep overlapping of electron wavefunctions to form the $s_\pm$ state $(e-\downarrow, e-\uparrow)_{HM}$ (see Figure 1). A common evidence to support this mechanism is the observation that for FeSe which is the simplest member of the iron-based compounds, its structure and magnetic properties depend sensitively on the ratio of Se:Fe [57]. For example its $T_c$ of 8 k can be pushed up to 14 K simply by replacing Se with Te. Since $Se^{2-}$ and $Te^{2-}$ have the same valence but different ionic radii, the substitution does not directly lead to charge carrier doping but to a new effective valence of the $3d^{z+}$.

In order to compare with experimental data for the iron based compounds as done for the cuprates, we now turn to a realization of the expression in Eq.(17b) in a similar form [25]:

$$T_C = 467.0 z \exp\left(-\frac{13.6}{z}\right) (^0 K) \qquad (19)$$

where 467.0 is the experimental Debye temperature of iron (see Table 1). It is also plotted alongside the result for the cuprates in Table 1. There is good agreement with the experimental data in the iron based compounds as shown in Table 1 and Figure 2.

**TABLE 1.** Dependence of $T_c$ on the effective valence z of $Cu^{z+}$ in the cuprates and $Fe^{z+}$ in the iron pnictides.

| z | $T_c(Cu^{z+})$ | $T_c(Fe^{z+})$ | z | $T_c(Cu^{z+})$ | $T_c(Fe^{z+})$ |
|---|---|---|---|---|---|
| 1.0 | 0.0005 | 0.0006 | 4.0 | 49.0320 | 62.2413 |
| 1.5 | 0.0636 | 0.0809 | 4.1 | 54.6032 | 69.4247 |
| 2.0 | 0.8182 | 1.0403 | 4.2 | 60.5318 | 76.9625 |
| 2.5 | 3.9847 | 5.0663 | 4.3 | 66.8201 | 84.9477 |
| 3.0 | 11.8397 | 15.0535 | 4.4 | 73.4698 | 93.4125 |
| 3.2 | 16.7656 | 21.3165 | 4.5 | 80.4820 | 102.328 |
| 3.4 | 22.8729 | 29.0816 | 5.0 | 120.9790 | 153.817 |
| 3.5 | 26.3964 | 33.5614 | 5.5 | 170.4087 | 216.664 |
| 3.6 | 30.2451 | 38.4549 | 6.0 | 228.4397 | 290.447 |
| 3.8 | 38.9482 | 49.5203 | | | |

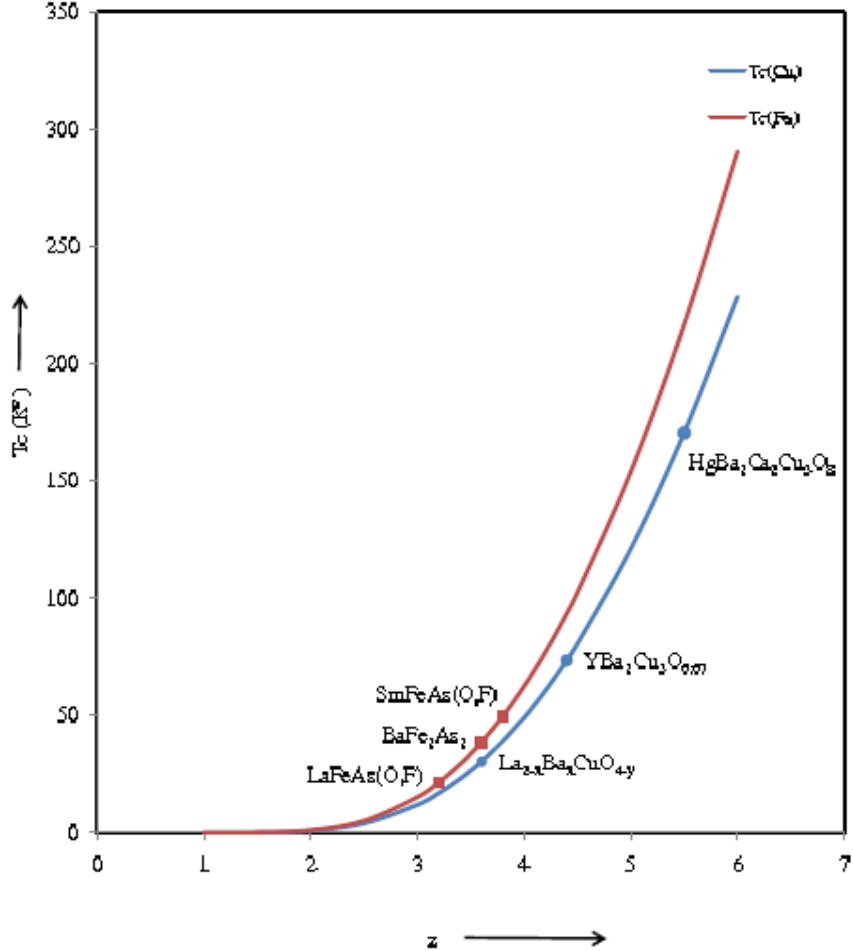

**FIGURE 2.** (Colour online) Predicted dependence of the transition temperature $T_c$ on the effective valence z for the cuprates in Eq.(18) and the iron based superconductors in Eq.(19)

## Summary and Conclusion

The absence of the electron-phonon interaction mechanism of the BCS theory in other areas of particle physics [29], its limitation to conventional superconductors and predictive power to 25 K [7] and a number of other flaws (see Ref. [6]) instigate the need to seek a more natural mechanism for the formation of the Cooper pair. The natural description is the foundation of the iso-superconductivity model, wherein the $Cu^{z+}$ ion of effective valence, $z \equiv 2(n - x)/m$ or an $Fe^{z+}$ ion of effective valence of appropriate value provides a "trigger" for the deep overlapping of electron wavefunctions to form the Cooper pair (e−↓, e−↑) for both the cuprates and iron based compounds respectively. An obvious open problem now is how to design real materials from the predictive values of the effective valence. This requires further investigation on how parameters from the periodic table database, electronegativity spectrum and material database can be used to obtain the effective valence of Cu and Fe as the control parameter for both the superconducting cuprates and iron based compounds [43]. It will also be interesting to investigate the effect of doping on the iso-superconductivity model for non-conventional superconductors. Speculatively, if the approach of assuming the Coulomb interaction as a screened interaction mediated by a Bosefield is to merely allow one to define different approximations as pointed out in Ref [26], such that the electron-phonon interaction of the BCS theory is a mathematical artifact as being canvassed by some workers [6, 8, 29], then the electron-phonon coupling constant will also be a mathematical artifact that has been simulating some material specific parameters of the effective valence. These are important investigations to be made on the foundation of iso-superconductivity.

## ACKNOWLEDGMENTS

We are grateful to the Santalli 's Foundation for its support of this work. GEA acknowledges that part of this work was done at the Max Planck Institute for Physics of Complex Systems, Dresden, Germany